\documentclass[10pt,letterpaper,twocolumn]{article} 

\usepackage{ol2}
\usepackage[draft]{hyperref}
\usepackage{amsmath}

\begin{document}

\twocolumn[ 

\title{Raman-induced limits to efficient squeezing in optical fibers}


\author{Ruifang Dong$^{1,*}$, Joel Heersink$^1$, Joel F. Corney$^2$, Peter D. Drummond$^2$, Ulrik L. Andersen$^{1,3}$, and Gerd Leuchs$^1$}

\address{
$^1$Institut f\"ur Optik, Information und Photonik, Max--Planck Forschungsgruppe, Universit\"at Erlangen--N\"urnberg, G\"{u}nther-Scharowsky-Str. 1, Bau 24, 91058, Erlangen, Germany \\
$^2$ARC Centre of Excellence for Quantum-Atom Optics, School of Physical Sciences, The University of Queensland, Brisbane, QLD 4072, Australia \\
$^3$Department of Physics, Technical University of Denmark, Building 309, 2800 Lyngby, Denmark \\
$^*$Corresponding author: rdong@optik.uni-erlangen.de
}

\begin{abstract}We report new experiments on polarization squeezing using ultrashort photonic pulses in a single pass of a birefringent fiber. We measure what is to our knowledge a record squeezing of $-6.8\pm 0.3$~dB in optical fibers which when corrected for linear losses is $-10.4\pm0.8$~dB. The measured polarization squeezing as a function of optical pulse energy, which spans a wide range from 3.5-178.8~pJ, shows a very good agreement with the quantum simulations and for the first time we see the experimental proof that Raman effects limit and reduce squeezing at high pulse energy. \end{abstract}


 ] 

\noindent Quantum communication and information science are undeniably important areas in modern physics, highlighted by their rapid expansion in recent years. Within the framework of quantum continuous variable information protocols~\cite{braunstein.book}, nonclassical polarization states of light have recently attracted particular interest ~\cite{chirkin93.qe,grangier87.prl,bowen02.prl,josse03.prl,heersink03.pra,heersink05.ol} due to their compatibility with the spin variables of atomic systems~\cite{hald01.jomo} and their simple detection without additional local oscillators~\cite{korolkova02.pra}. The early experimental generation of polarization squeezing used continuous wave light and parametric processes~\cite{grangier87.prl,bowen02.prl,hald01.jomo}. Since these experiments, polarization squeezing has also been experimentally demonstrated using cold atomic samples~\cite{josse03.prl} and silica fibers\cite{heersink03.pra}.

A significant improvement in the fiber based production of polarization squeezing came with the implementation of the single-pass method~\cite{heersink05.ol}. This scheme was greatly simplified compared with previous experiments and led to the possibility of squeezing at any given pulse energy. Based on the Kerr effect in an optical fiber, two equally strong ultrashort pulses co-propagated along the fiber's two orthogonal polarization axes and produced two quadrature-squeezed beams with statistically identical uncertainties. The pulses were combined in a Stokes measurement to give polarization squeezing up to $-5.1\pm0.3$~dB. This experiment also showed potential for further improvement of the squeezing by optimizing the losses in the setup. Subsequent first-principles simulations showed good agreement with the experimental measurements~\cite{corney06.prl} and predicted Raman-induced limits on the squeezing at energies beyond what the experiment had investigated.
 
In this letter we present new results of polarization squeezing based on an optimized version of the single-pass scheme. Furthermore, a substantially wide range of pulse energies is investigated to fully characterize the quantum noise property of the ultrashort pulses in the fiber, the comparisons of the experimental measurements with the simulations are also demonstrated and the very good agreement reveals the evidence that the deteriorated squeezing at high energy is due to the Raman-effects. 

\begin{figure}[htb]
  \centerline{
    \includegraphics[width = 8.1cm]{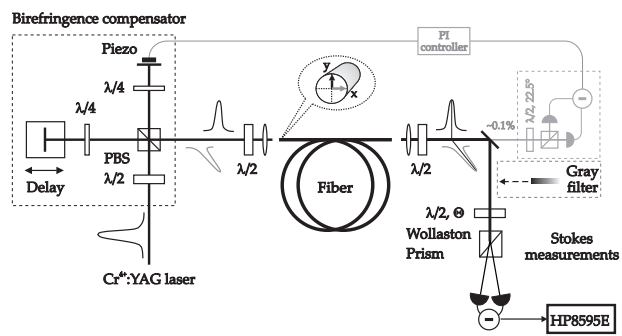}
	}
	\caption{Experimental setup for efficient polarization squeezing generation. PBS, polarizing beam splitter; $(\lambda /2)$, half-wave plates; $(\lambda /4)$, quarter-wave plates.}
	\label{fig:setup}
\end{figure}

The experimental setup is shown in Fig.~\ref{fig:setup}. A home-made Cr$^{4+}$:YAG laser is used as a source of 140~fs (FWHM) sech-shaped optical pulses at a repetition rate of 163~MHz. The spectra are centered at $\lambda_{0}=1499.5~nm$ with a bandwidth of 19~nm. The fiber we use is a 13.2~m-long polarization-maintaining (PM) fiber, 3M FS-PM-7811 with a 5.7~$\mu$m core diameter and an average nonlinear refractive index of $n_{2}=2.9\times10^{-20}~m^{2}/W$. The second-order and third-order dispersion of the fiber at the optical wavelength $\lambda_{0}$ are experimentally determined to be $\beta_{2}=-11.1~fs^{2}/mm$ and $\beta_{3}=83.8~fs^{3}/mm$ respectively using white-light interferometry.  

We prepare ultrashort pulses of the same optical power and couple them into the two orthogonal polarization axes, $x$ and $y$, of the fiber. After the fiber, we obtain two independent Kerr squeezed beams with approximately the same quadrature noise property $\Delta^{2} \hat{X}_{x,\theta}=\Delta^{2} \hat{X}_{y,\theta}=\Delta^{2} \hat{X}_{\theta}$ (the squeezed quadratures rotated by an angle $\theta_{sq}$ relative to the amplitude quadrature). To generate the polarization squeezing, the two emerging pulses were temporally overlapped with a $\pi/2$ relative phase shift by use of the birefringence precompensation with an unbalanced Michelson-like interferometer~\cite{heersink03.pra,heersink05.ol,fiorentino01.pra} as well as the locking loop based on 0.1\% of the fiber output. The resultant beam is thus circularly polarized ($\hat{S}_{3}\neq0$). The corresponding Heisenberg uncertainty relations are reduced to a single nontrivial one for the Stokes parameters in the so-called $\hat{S_{1}}-\hat{S_{2}}$ dark plane, denoted by $\Delta^{2} \hat{S}_{\theta}\, \Delta^{2} \hat{S}_{\theta+\pi/2}\geq{}|\langle\hat{S}_3\rangle|^{2}$, and the polarization squeezing is linked to quadrature squeezing by $\Delta^{2}\hat{S}_{\theta_{sq}} = \alpha^2 \Delta^2 \hat{X}_{\theta_{sq}}$~\cite{heersink05.ol}. 

The noise of a given Stokes parameter is measured with the help of a half-wave plate $(\lambda /2)$ and a polarizing beam splitter. We use a Wollaston prism for higher extinction ratio ($>10^{-5}$ for both polarizations) and the outputs are detected directly by use of a pair of balanced photo-detectors based on custom-made pin photo-diodes (Laser Components GmbH, 98\% quantum efficiency at DC). The difference of the detected AC photocurrents provides measurement of the optical noise. The shot noise limit is determined by use of the fiber-bypass beam from the laser that was verified to be shot-noise limited by balanced detection. The measurement is recorded by a HP8595E spectrum analyzer operated at 17.5~MHz with 300~kHz resolution bandwidth and 30 Hz video bandwidth, the measured noise traces are corrected for an electronic noise of $-85.1$~dBm. 

\begin{figure}[htb]
  \centerline{
    \includegraphics[width = 5cm]{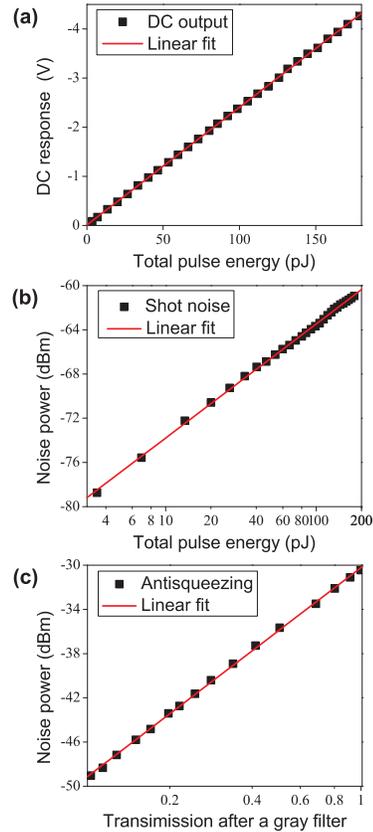}
	}
	\caption{Plots of (a) DC response of one of the detectors pair and (b) noise power measured at 17.5 MHz by the minus output of the detectors pair versus the optical pulse energy from the fiber. (c) The measured anti-squeezed noise as a function of the transmittance after a gray filter.}
	\label{fig:saturetest}
\end{figure}

To ensure that the balanced detectors are not saturated by the strong optical powers in the experiments, we establish a set of measurements to check both the DC and AC outputs as a function of the optical pulse energy from 3.5-178.8~pJ. In Fig.~\ref{fig:saturetest}(a) and (b) we show the DC response of one single detector as well as the difference of the
AC photocurrents from the pair of detectors versus the optical energy of a coherent state. From the plot it can be seen that the detector response is linear, and therefore that the squeezing measurements carried out in this region are unsaturated. 

However, saturation is still possible for the antisqueezing measurement as the RF noise for a given optical power is much greater. We check this using a variable gray filter with which the optical power of the squeezed beam is varied, as shown in Fig.~\ref{fig:setup}. The optical pulse energy was fixed to the maximum (178.8~pJ) and the Stokes measurement was set to measure the anti-squeezed quadrature. The measurement of the noise variance as a function of the transmittance of the optical field through the gray filter is shown in Fig.~\ref{fig:saturetest}(c). The linear dependence confirms that the detectors are unsaturated even up to the maximum optical power achievable.

\begin{figure}[htb]
  \centerline{
    \includegraphics[width = 5cm]{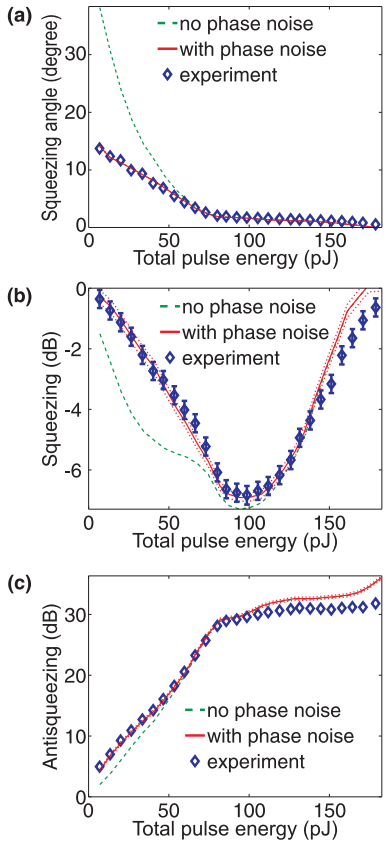}
	}
	\caption{Diamonds show the experimental results of the (a) squeezing angle, (b) squeezing and (c) antisqueezing for the 13.2~m fiber. Error bars on the squeezing data indicate the uncertainty in the noise measurement; for the antisqueezing, the error bars were too small to be plotted. Solid and dashed lines show the simulation results with and without additional phase noise respectively. Dotted lines indicate the sampling error in the simulation results.}
	\label{fig:results_13.2m}
\end{figure}

The squeezed and anti-squeezed quadratures and the squeezing angle, $\theta_{sq}$ were experimentally investigated as a function of the pulse energy from 3.5~pJ to 178.8~pJ and the results are plotted in Fig.~\ref{fig:results_13.2m}. The maximum observed squeezing is $-6.8\pm0.3$~dB at an energy of 98.6~pJ. The corresponding antisqueezing of this state is $29.6\pm0.3$~dB and the squeezing angle is $1.71^{o}$. The loss of the setup was found to be 13\%: 5\% from the fiber end, 4.6\% from optical elements, attenuation of the fiber (2.03~dB/km@1550~nm), 2\% from incomplete interference between two polarization modes (99\% visibility was measured) and 2\% from the photodiodes. Thus we infer a maximum polarization squeezing of $-10.4\pm0.8$~dB. As the optical energy goes beyond 98.6~pJ, the squeezing is reduced, eventually reaching the shot noise limit (SNL), and the increment of antisqueezing slows down to a plateau area. By applying the first-principles quantum dynamics of radiation propagating in a single-mode optical fiber and phase-space methods~\cite{corney06.prl,drummond01.josab} which includes the effects of dispersions up to the third order and the $\chi^{(3)}$ nonlinearity as well as the Raman coupling to thermal phonons, the squeezing, antisqueezing and squeezing angle at different input energies are simulated. The fraction of the nonlinearity that is due to the Raman gain is estimated as 15\%, and the photon number ($2\bar{n}$) in a fundamental soliton pulse as $4.5\times10^{8}$. The excess phase noise, such as depolarizing guided acoustic wave Brillouin scattering (GAWBS)~\cite{corney06.prl}, is estimated by fitting the simulated squeezing angles to the experimentally measured squeezing angles as shown by red solid line in Fig.~\ref{fig:results_13.2m}(a). After taking the 13\% linear loss into account, the theoretical results for squeezing and antisqueezing which are given in Fig.~\ref{fig:results_13.2m}(b) and (c) by red solid lines achieve a very good match with the experimental results. From the simulations, the effect of the GAWBS is seen to be a reduction in squeezing for lower pulse energies. Above the soliton energy ($\approx$120~pJ), the deterioration of squeezing is attributed to the Raman effects since it can't be reasoned by the simulations with only electronic nonlinearity and dispersive effects; in addition, the third-order dispersion (TOD) has a noticeable effect on the squeezing.  

In conclusion, we have demonstrated what is to our knowledge a new record of squeezing $-6.8\pm0.3$~dB ($-10.4\pm0.8$~dB after correction for linear losses) in optical fibers. We also provided the experimental evidence of the Raman-induced limit to the squeezing in the fiber, which fits very well with the simulations. Although there is still residual discrepancy between simulations and experiments in the squeezed and antisqueezed quadratures at higher energy, which could be due to effects such as imperfect Raman spectrum modelling or initial pulse-shape distortion, the theoretical model has been proved to be comprehensive and highly efficient. Through the simulation work, the further improvement of squeezing generated in fibers is expected by estimating the optimal fiber length and the best trade-off in terms of squeezing and antisqueezing noise quantitatively. This highly squeezed polarization state is also a good source for applications in quantum information and communication, particularly due to the ease of detection without the need for an external phase reference, for example, distillation of quantum states afflicted by non-Gaussian noise~\cite{heersink06.prl} and entangled states~\cite{dong07.prep}.

This work was funded by EU Project COVAQIAL. J. F. Corney and P. D. Drummond are supported by the Australian Research Council. 


\begin{thebibliography}{10}

\bibitem{braunstein.book} S. L. Braunstein and A. K. Pati, eds., {\it Quantum information theory with continuous variables.} (Kluwer, Dordrecht, 2002).

\bibitem{chirkin93.qe} A. S. Chirkin, A. A. Orlov and D. Yu. Paraschuk, Kvantovay Elektronika {\bf 20}, 999 (1993). Quantum Elec. {\bf  23}, 870-874 (1993). 

\bibitem{grangier87.prl} P. Grangier, R. E. Slusher, B. Yurke and A. LaPorta, Phys. Rev. Lett. {\bf 59}, 2153-2156 (1987). 

\bibitem{bowen02.prl} W. P. Bowen, R. Schnabel, H. A. Bachor and P.K. Lam, Phys. Rev. Lett. {\bf 88}, 093601-093604 (2002). 

\bibitem{josse03.prl} V. Josse, A. Dantan, L. Vernac, A. Bramati, M. Pinard and E. Giacobino, Phys. Rev. Lett. {\bf 91}, 103601-103604 (2003).

\bibitem{heersink03.pra} J. Heersink, T. Gaber, S. Lorenz, O. Gl\"ockl, N. Korolkova and G. Leuchs, Phys. Rev. A {\bf 68}, 013815-013827 (2003).

\bibitem{heersink05.ol} J. Heersink, V. Josse, G. Leuchs, and U. L. Andersen, Opt. Lett. {\bf 30}, 1092-1094 (2005).

\bibitem{hald01.jomo} J. Hald, J. L. Sorensen, C. Schori and E. S. Polzik, J. Mod. Opt. {\bf 47}, 2599-2614 (2001).

\bibitem{korolkova02.pra} N. Korolkova, G. Leuchs, R. Loudon, T. C. Ralph and Ch. Silberhorn, Phys. Rev. A {\bf 65}, 052306-052317 (2002).

\bibitem{corney06.prl} J. F. Corney, P. D. Drummond, J. Heersink, V. Josse, G. Leuchs, and U. L. Andersen, Phys. Rev. Lett. {\bf 97}, 023606 (2006). 

\bibitem{fiorentino01.pra} M. Fiorentino, J. E. Sharping, P. Kumar, D. Levandovsky and M. Vasilyev, Phys. Rev. A, {\bf 64}, 031801(R)-031804 (2001).

\bibitem{drummond01.josab} P. D. Drummond and J. F. Corney, J. Opt. Soc. Am. B {\bf 18}, 139152 (2001).

\bibitem{heersink06.prl} J. Heersink, Ch. Marquardt, R. F. Dong, R. Filip, S. Lorenz, G. Leuchs, and U.L. Andersen, Phys. Rev. Lett. {\bf 96}, 253601 (2006).

\bibitem{dong07.prep} R. F. Dong, J. Heersink, J. Yoshikawa, O. Gl\"ockl, U.-L. Andersen, and G. Leuchs, arXiv:quant-ph/0709.2237 (2007).

\end{thebibliography}
\end{document}